\documentclass[aps,prl,twocolumn,groupedaddress]{revtex4}
\usepackage{graphicx}

\begin{document}

\title{Quantum Lifetime of Two-Dimensional Holes}

\author{J.~P. Eisenstein$^1$, D. Syphers$^{1,2}$, L.~N. Pfeiffer$^3$, and K.~W. West$^3$}

\affiliation{$^1$California Institute of Technology, Pasadena CA 91125 
\\$^2$Bowdoin College, Brunswick, ME 04011\\
	 $^3$Bell Laboratories, Alcatel-Lucent Inc., Murray Hill, NJ 07974\\}

\date{\today}

\begin{abstract}The quantum lifetime of two-dimensional holes in a GaAs/AlGaAs double quantum well is determined via tunneling spectroscopy.  At low temperatures the lifetime is limited by impurity scattering but at higher temperatures hole-hole Coulomb scattering dominates.  Our results are consistent with Fermi liquid theory, at least up to $r_s = 11$.  At the highest temperatures the measured width of the hole spectral function becomes comparable to the Fermi energy.  A new, tunneling-spectroscopic, method for determining the in-plane effective mass of the holes is also demonstrated.
\end{abstract}


\maketitle
In Landau's Fermi liquid theory\cite{pines} the strongly interacting many-fermion system is modeled as an ensemble of weakly-interacting quasiparticles experiencing an infinite set of ``molecular'' mean fields.   Central to this theory is the notion that at low temperatures and close to the Fermi surface the quasiparticle lifetime is long. For example, in a disorder-free two-dimensional system the lifetime $\tau$ of a quasiparticle at the Fermi level diverges as $\tau^{-1} \sim T^2ln(T_F/T)$ at temperatures $T$ low compared to the Fermi temperature $T_F$\cite{hodges}.  

Fermi liquid theory is expected to fail at sufficiently low density $n$ where Coulomb interactions overwhelm kinetic effects and the electron liquid freezes into a Wigner crystal. At $T = 0$ this transition is expected to occur near $r_s \approx 37$ in two dimensions, where $r_s=a_0^{-1}(\pi n)^{-1/2}$ is the Wigner-Seitz radius and $a_0$ the relevant Bohr radius\cite{ceperley,spivak}. For 2D electrons in GaAs with their low effective mass ($m^*/m_0 = 0.067$) this corresponds to the extremely low density of $2.3 \times 10^8$ cm$^{-2}$.  For this reason, the strongly correlated, high $r_s$ regime is often pursued using 2D {\it holes} in GaAs.  The higher mass of the holes ($m^*/m_0 \sim 0.2$ to 0.5) reduces the Bohr radius and thus provides a larger $r_s$ than a 2D electron system of the same density.  For these reasons determining whether the quantum lifetime of 2D holes obeys the Fermi liquid expectation is both desirable and timely.

Conventional transport measurements are an effective tool for determining the carrier lifetime when it is limited by impurity or phonon scattering.  They are, however, ineffective when carrier-carrier Coulomb scattering dominates because such collisions conserve the net momentum of the system and thus do not show up directly in the resistivity.  In contrast, tunneling provides access to the underlying spectral functions of the system and thus the total lifetime broadening of the quasiparticle states.  This access is especially direct in the case of tunneling between parallel 2D systems since the conservation of in-plane momentum enforces conservation of kinetic energy as well.  Previous experiments\cite{murphy} on tunneling between parallel 2D electron systems have clearly demonstrated the ability to measure the quantum lifetime of the electrons.  In the regime dominated by electron-electron scattering these measurements were subsequently found to be in quantitative agreement with theoretical calculations of the quantum lifetime which included local-field corrections to the random phase approximation\cite{jungwirth,zheng}.  In this paper we present an analogous tunneling study of the quantum lifetime of 2D holes in GaAs quantum wells.  Our results demonstrate that, at least for $r_s \leq 11$, the quantum lifetime of 2D holes is quite consistent with Fermi liquid theory, although the sophisticated calculations needed for truly quantitative comparison have not been attempted.  We remark that at high temperatures and low density the observed lifetime broadening of the holes becomes comparable to the Fermi energy itself.

The sample used in this experiment is a conventional GaAs/AlGaAs double quantum well (DQW) heterostructure grown by molecular beam epitaxy on a $\langle 001 \rangle$-oriented GaAs substrate.  Two 15 nm GaAs quantum wells are separated by a 14.5 nm Al$_{0.3}$Ga$_{0.7}$As barrier layer.  Carbon doping sheets placed 80 nm above and below the quantum wells in Al$_{0.3}$Ga$_{0.7}$As cladding layers populate the quantum wells with 2D holes.  The as-grown density $p$ and low temperature mobility $\mu$ of the 2DHS in each well are approximately $7\times10^{10}$ cm$^{-2}$ and $7.5\times10^5$ cm$^2$/Vs, respectively. At these low densities only the lowest heavy-hole subband of the wells is occupied.  The sample is patterned into a square mesa, 250 $\mu$m on a side, with 40 $\mu$m-wide arms extending outward to diffused InZn ohmic contacts. A well-established selective depletion scheme is used to allow independent electrical contact to either 2D layer separately\cite{sepcon}.  Aluminum Schottky gates deposited on the front and backside of the thinned sample allow the density in the two layers in the central mesa region to be varied independently over a wide range.  Interlayer tunneling conductance measurements are made by applying a small (80 $\mu$V, typically), low frequency (13 Hz) voltage $dV$ between the two layers and measuring the resulting ac current $dI$ which flows. Adding a dc interlayer bias $V$ allows the $dI/dV$ vs. $V$ tunneling conductance spectrum to be recorded. Here we confine our attention to tunneling between the ground subbands of two quantum wells.

\begin{figure}
\centering
\includegraphics[width=3.2 in, bb=0 0 263 180]{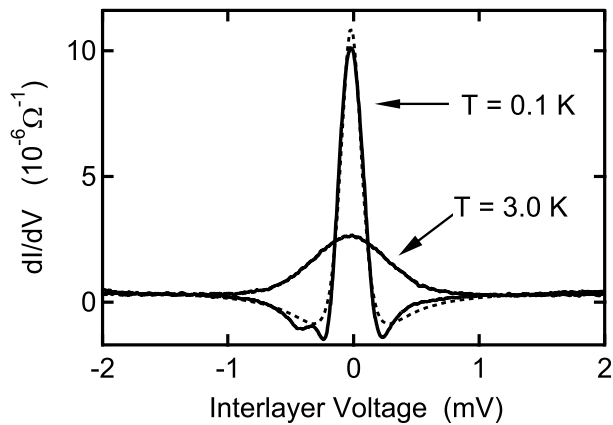}
\caption{\label{fig:fig1}Typical tunneling conductance resonances in a double layer 2D hole system.  Dashed line represents fit to lorentzian spectral function described in text.  (At $T = 3$ K the fit is almost indistinguishable from the data.)}
\end{figure}
Figure 1 shows two typical tunneling conductance spectra.  These data, taken at $T$ = 0.1 and 3.0 K, illustrate the sharply resonant character of tunneling between parallel 2D systems.  Conservation of energy and in-plane momentum drastically reduces the phase space for 2D-2D tunneling. Indeed, tunneling between identical and ideal parallel 2D hole (or electron) systems, can only occur when the quantum well subbands line up precisely. This is true even if the subband dispersion in the wells is non-parabolic, provided it is identically so in the two layers. With equal densities in the layers (a $balanced$ DQW) the alignment of the subband edges occurs simultaneously with the alignment of the Fermi levels and thus the resonance is centered at zero interlayer voltage. (In a density imbalanced DQW the resonance shifts to finite interlayer voltage; we shall return to this case later.) For the data in Fig. 1 each 2D hole gas has a density of $p_1 = p_2 = 7.2 \times 10^{10}$ cm$^{-2}$\cite{density}. 

As Fig. 1 illustrates, the width of the tunneling conductance resonance grows with increasing temperature. This is not due to simple thermal smearing of the Fermi distributions of the two 2DHSs.  The twin constraints of energy and momentum conservation require precise subband alignment for tunneling to occur, irrespective of the carrier distribution function, thermal or otherwise.  Instead, the width of the tunnel resonances stems from the finite lifetime of the momentum states themselves.  As demonstrated previously, measurements of tunneling between parallel 2D electron gases in GaAs yield determinations of the lifetime of 2D electrons due to Coulomb scattering which are in quantitative agreement with theory\cite{murphy,jungwirth,zheng}.  Following this earlier work, we shall initially assume that the 2D heavy hole bands are parabolic, axially symmetric, and spin-degenerate, and that the hole spectral function $A(E,{\bf k})$ is well-approximated by a function $A(x)$ of the single variable\cite{energy} $x = E+E_F-\hbar^2k^2/2m$ with a sharp (relative to the Fermi energy $E_F$) ``quasiparticle'' peak near $x = 0$.  Under these assumptions the ratio $F(V)=I/V$ of tunnel current to interlayer voltage is just the convolution of the spectral functions in the ``left'' and ``right'' 2D layers:  
\begin{equation}
F(V) = \beta |t|^2
\!\! \int_{- \infty}^{\infty}\!\!\! dx A_L (x)
A_R (x+E_{F,R} -E_{F,L} -eV)
\end{equation}
with $|t|$ the tunneling matrix element and $\beta$ a constant.  Note that no Fermi distribution functions appear in this equation; temperature dependence enters only through the spectral functions. For quasiparticles with quantum lifetime $\tau$, the spectral function is a lorentzian: $A(x)=(\gamma/ \pi )/(x^2 + \gamma^2)$, with $\gamma = \hbar / 2\tau$. The ratio $F(V)$ is therefore also a lorentzian, centered at $eV = E_{F,R}-E_{F,L}$ (i.e. when the subband edges in the two quantum wells are aligned) and possessing a half-width at half maximum of $\Gamma = \gamma_L +\gamma_R$. Thus $\Gamma$ is the sum of the lifetime broadening of the holes in the two 2D layers. 

\begin{figure}
\centering
\includegraphics[width=3.2 in, bb=0 0 266 183]{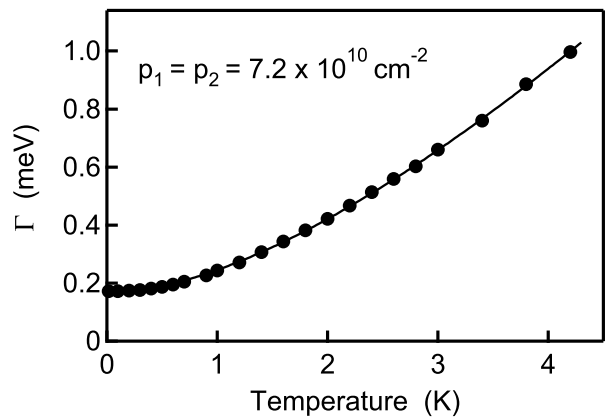}
\caption{\label{fig:fig2}Temperature dependence of tunneling linewidth $\Gamma$ extracted from fitting procedure described in text. Solid line is a guide to the eye.} 
\end{figure}
Assuming $F(V)$ is a lorentzian, the broadening parameter $\Gamma$ has been extracted from our tunneling $dI/dV$ data by fitting them to $G(V) \equiv F+VdF/dV +\alpha$.  (The constant $\alpha$ is added to account for the small non-resonant background conductance that we observe. At low temperatures $\alpha$ is typically less than 5-10\% of the peak conductance on resonance.) As Fig. 1 demonstrates, these fits are quite good at the relatively high density of $7.2 \times 10^{10}$ cm$^{-2}$ per layer, especially at the higher temperatures studied ($T \leq 4.2$ K for all data in this paper.)  At lower densities the fits remain good at low and intermediate temperatures where the linewidth is relatively small but degrade somewhat at higher temperatures as the broadening becomes comparable to the Fermi energy.

Figure 2 shows the temperature variation of the linewidth $\Gamma$ for $p_1 = p_2 = 7.2 \times 10^{10}$ cm$^{-2}$.  Below about $T = 0.5$ K the linewidth is temperature independent.  This is not surprising since inelastic processes (hole-hole and hole-phonon scattering) vanish at low temperatures and the temperature independent elastic scattering of holes off of the static disorder potential in the sample dominates.  For the data in Fig. 2 the low temperature limiting linewidth is approximately $\Gamma = 0.17$ meV.  While this number is large compared to that which the high sample mobility would suggest, it compares very favorably with the quantum lifetime deduced from the onset of Shubnikov-de Haas oscillations in the tunneling conductance\cite{sdh}. It has long been known that the low temperature mobility greatly over-estimates the quantum lifetime of carriers in modulation-doped systems owing to the predominance of small angle scattering\cite{smallangle}.

At higher temperatures the linewidth begins to increase steadily. Hole-phonon scattering is not a major contributor to this increase.  Since the 2DHS mobility falls by only a factor of $\sim 2.5$ upon warming to $T = 4.2$ K, we conclude that the hole-phonon scattering rate $\tau_{hp}^{-1}$ does not significantly exceed the limiting low temperature mobility scattering rate $\tau_{\mu}^{-1}$ over the entire temperature range of these experiments. Since $\tau_{\mu}^{-1}$ is already much too small to explain our tunneling linewidths\cite{sdh}, so must be $\tau_{hp}^{-1}$\cite{phonons}.  We instead attribute the increasing tunneling linewidth in our temperature range primarily to hole-hole Coulomb scattering processes. This is the same conclusion reached for the case of tunneling between parallel 2D electron gases\cite{murphy}.

In order to compare our findings with theoretical estimates of the hole-hole scattering rate, we must first estimate the Fermi energies of the 2DHS or, if we continue to assume the hole dispersion is parabolic, the in-plane hole effective mass, $m^*$. The valence band structure in GaAs is sufficiently complicated that the value of the effective mass depends substantially on the details of the potential used to confine the holes.  For example, it was long ago shown that spin-orbit contributions to the hole dispersions depend qualitatively on the presence or lack of inversion symmetry of the confinement potential\cite{eisenstein84}. Very recently Zhu, {\it et al.}\cite{zhu} have performed a cyclotron resonance study of hole effective masses in GaAs quantum wells and find a significant dependence of $m^*$ upon the width of the wells and the density of the 2DHS confined within them. Here we demonstrate a new way to determine the effective mass at zero magnetic field. More precisely, we employ tunneling to measure the derivative $dE_F/dp$ of the Fermi energy with respect to the carrier concentration. The effective mass at the Fermi level then follows: $m^* = \pi\hbar^2(dE_F/dp)^{-1}$.  
\begin{figure}
\centering
\includegraphics[width=3.2 in, bb=131 122 450 310]{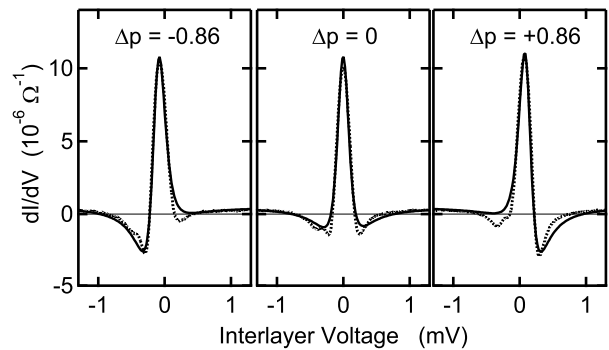}
\caption{\label{fig:fig3} Tunneling spectra (dashed lines) at $T = 0.1$ K for balanced (center panel) and slightly imbalanced double layer 2D hole systems. Solid lines are fits described in text. The density differences, $\Delta p$, are in units of $10^{10}$ cm$^{-2}$.} 
\end{figure}

As discussed above, tunneling $dI/dV$ resonances are centered at $V = 0$ only when the densities in the two layers are equal.  If the densities are unequal, the conductance resonances shift to finite voltage and take on a derivative shape.  The observed $dI/dV$ data may be analyzed in the same way as before, only now allowing the underlying lorentzian spectral convolution $F(V)$ to be centered at non-zero voltage, as Eq. 1 suggests.  The best-fit voltage shift $V_s$ then gives the difference in Fermi energies of the two layers: $eV_s = E_{F,R}-E_{F,L}$. Figure 3 illustrates this approach with three tunnel resonances, one with balanced layers (with $p_1 = p_2 = 7.2 \times 10^{10}$ cm$^{-2}$) and two others in which a small imbalance has been induced\cite{hall}.  For the data shown in the figure, the density differences between the two layers are equal but opposite in sign: $\Delta p =p_1$ - $p_2 = \pm 0.86 \times 10^{10}$ cm$^{-2}$. 

Figure 3 clearly demonstrates that the resonances shift and distort anti-symmetrically with the applied density difference between the layers. The dashed lines are the actual tunneling conductance data and the solid lines are the fits to a shifted lorentzian spectral convolution $F(V)$.  For the data shown the best-fit voltage shifts of the lorentzian $F(V)$ are $eV_s = +0.115$ and $-0.110$ meV for $\Delta p = +0.86$ and $-0.86 \times 10^{10}$ cm$^{-2}$.  This implies $m^*/m_0 \approx 0.18$, with $m_0$ the mass of an electron in vacuum. With this effective mass the Fermi energy/temperature of a 2DHS with density $7.2 \times 10^{10}$ cm$^{-2}$ is $E_F = 0.95$ meV, or $T_F =11$ K, assuming the bands are parabolic and spin-degenerate.  The Wigner-Seitz radius for this density and mass is $r_s = 5.7$.

Using this same tunneling approach we have measured the temperature dependent lifetime broadening $\Gamma$ and estimated the effective mass, Fermi temperatures, and $r_s$ values in this same sample at two lower densities: $p_1=p_2=4.6$ and $3.9 \times 10^{10}$ cm$^{-2}$.  We estimate $m^*/m_0 \approx 0.24$, $T_F=5.3$ K, and $r_s = 9.5$ at the higher density and $m^*/m_0 \approx 0.26$, $T_F= 4.2$ K, and $r_s = 11.1$ at the lower.  These effective masses are comparable to those recently reported by Zhu, {\it et al.}\cite{zhu}.  

With these Fermi energies in hand we can normalize the lifetime broadening $\Gamma$ and temperature $T$.  The results of this normalization are presented in Fig. 4.  The solid dots represent our results for $p_1=p_2=7.2 \times 10^{10}$ cm$^{-2}$, the open dots for $p_1=p_2=4.6 \times 10^{10}$ cm$^{-2}$ and finally the solid triangles for $p_1=p_2=3.9\times 10^{10}$ cm$^{-2}$.  In each case we have subtracted $\Gamma_0 \approx 0.17$ meV, the low temperature limiting value of the linewidth, in order to concentrate on the inelastic, hole-hole scattering contribution\cite{gamma0}.  
\begin{figure}
\centering
\includegraphics[width=3.2 in, bb=0 0 265 183]{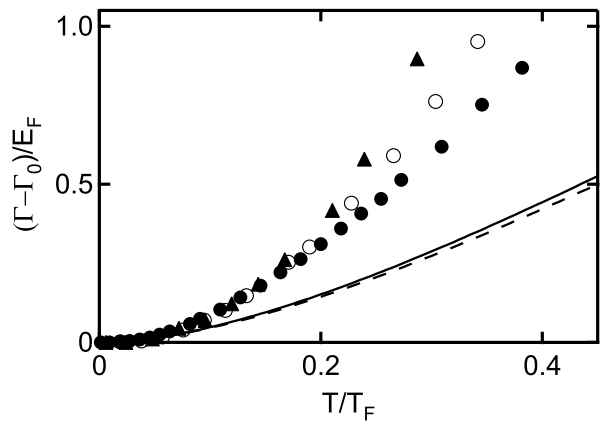}
\caption{\label{fig:fig4}Normalized tunneling linewidth, with low temperature limit subtracted, versus $T/T_F$ for three different densities: $7.2 \times 10^{10}$ cm$^{-2}$ (solid dots); $4.6 \times 10^{10}$ cm$^{-2}$ (open circles); and $3.9 \times 10^{10}$ cm$^{-2}$ (triangles), per layer.  Solid and dashed lines are theoretical results which neglect off-shell energy dependences and local field corrections to the RPA.} 
\end{figure}

For $T/T_F < 0.2$ Fig. 4 reveals that the normalized inelastic lifetime broadening is independent of density.  This is the result originally anticipated by Hodges, Smith and Wilkins\cite{hodges,jungwirth2}.  At higher temperature, where the linewidth is becoming comparable to $E_F$ (keeping in mind that $\Gamma$ is the {\it sum} of the lifetime broadening in the two layers), the broadening at low density begins to exceed that at high density.  The lines in the figure are analytical results taken from Ref.\cite{jungwirth}, the solid line appropriate to $r_s = 11.1$ (our lowest density) and the dashed line $r_s = 5.7$ (our highest density).  These theoretical curves, which reflect the RPA and are only appropriate for $T/T_F << 1$, include only the thermal lifetime broadening of quasiparticles at the Fermi level.  They do not incorporate the additional broadening arising from the energy dependence of the quasiparticle lifetime, nor any higher-order, local field corrections to the RPA.  Both effects were previously found to be important for achieving quantitative understanding of the analogous experimental results on 2D electron systems\cite{jungwirth,zheng,murphy}, and it seems reasonable to assume they will play a similar role here. 

Finally, we comment briefly on our neglect of non-parabolicity and spin-orbit splitting on the hole band structure.  While neither assumption is an excellent one, it is important to note two points.  First, the sharply resonant character of the tunneling conductance does not depend on either assumption, provided that both effects are the same in the two layers.  This is certainly the case for balanced quantum wells and will be nearly true for small imbalances and small interlayer voltages.  At large imbalance and large interlayer voltage the band structure in the two wells will begin to differ.  Indeed, this effect is observable in tunnel spectra taken under such conditions since they then fail to be well approximated by the simple lorentzian spectral functions we have assumed.  Second, the low 2D density in our experiments mitigates both non-parabolicity and spin-orbit splitting since both effects grow in importance with density.

In conclusion, via tunneling spectroscopy we have measured the quantum lifetime of 2D holes in GaAs quantum wells at low density.  Our results are consistent with the expectations of Fermi liquid theory, at least up to $r_s = 11$.

This work was supported by the NSF under Grant No. DMR-0242946 and the DOE under Grant No. DE-FG03-99ER45766.


\begin{thebibliography}{99}

\bibitem{pines} See, for example, {\it The Theory of Quantum Liquids, Vol. I} by D. Pines and P. Nozieres, (Benjamin, New York, 1966)

\bibitem{hodges} C. Hodges, H. Smith and J.W. Wilkins, Phys. Rev. B {\bf 4}, 302 (1971).

\bibitem{ceperley} B. Tanatar and D. Ceperley, Phys. Rev. B {\bf39}, 5005 (1989).

\bibitem{spivak} There have been recent suggestions of new phases between the electron liquid and Wigner crystal; see B. Spivak and S.A. Kivelson, Phys. Rev. B {\bf70}, 155114 (2004).

\bibitem{murphy} S.Q. Murphy, J.P. Eisenstein, L.N. Pfeiffer, and K.W. West, Phys. Rev. B {\bf 52}, 14825 (1995).

\bibitem{jungwirth} T. Jungwirth and A.H. MacDonald, Phys. Rev. B {\bf 53}, 7403 (1996).

\bibitem{zheng} L. Zheng and S. Das Sarma, Phys. Rev. B {\bf 53}, 9964 (1996).

\bibitem{sepcon} J.P. Eisenstein, L.N. Pfeiffer, and K.W. West, Appl. Phys. Lett. {\bf57}, 2324 (1990). 

\bibitem{density} The density is determined by observing the Shubnikov-de Haas-like oscillations of the peak tunnel conductance induced by a perpendicular magnetic field.

\bibitem{energy} The energy $E$ is measured relative to the Fermi level, $E_F$.


\bibitem{sdh} Assuming a hole effective mass of $m^*/m_0 = 0.2$, the transport lifetime deduced from the mobility suggests $\hbar/\tau_{\mu} = 0.008$ meV.  In contrast, the $B \approx 0.25$ T onset of magneto-oscillations of the tunneling conductance suggests $\hbar/\tau \approx 0.13$ meV, close to the low temperature limiting value of the tunneling spectral width $\Gamma$.  

\bibitem{smallangle}  S. Das Sarma and F. Stern, Phys. Rev. B {\bf 32}, 8442 (1985).

\bibitem{phonons} By $T = 4$ K the average acoustic phonon wavevector exceeds the diameter of the 2DHS Fermi sea.  Thus, unlike elastic scattering off of the static disorder potential, inelastic hole-phonon scattering events occur at all angles.


\bibitem{eisenstein84} J.P. Eisenstein, H.L. Stormer, V. Narayanamurti, A.C. Gossard, and W. Wiegmann, Phys. Rev. Lett. {\bf 53}, 2579 (1984).

\bibitem{zhu} H. Zhu, {\it et al.}, Solid State Commun. {\bf 141}, 510 (2007).

\bibitem{hall} Imbalance was created using the gate on the back surface of the sample.  The induced density change was thus primarily in the lower 2DHS layer, although finite compressibility effects led to a smaller change in the top layer as well.  The individual 2DHS densities are readily measured via the Hall effect in each layer.

\bibitem{gamma0} At low temperatures the tunneling linewidth was found to be independent of density.

\bibitem{jungwirth2} Ref. \cite{jungwirth} suggests that this is true only when $r_s$ is much less, or much greater, than unity. 



\end{thebibliography}
\end{document}